\documentclass[letterpaper]{article} 
\usepackage{aaai24}  
\usepackage{times}  
\usepackage{helvet}  
\usepackage{courier}  
\usepackage[hyphens]{url}  
\usepackage{graphicx} 
\usepackage{natbib}  
\usepackage{caption} 
\usepackage{algorithm}
\usepackage{algorithmic}
\usepackage{booktabs}
\usepackage{amsmath}
\usepackage{amssymb}
\usepackage{multirow}
\usepackage{bbding}
\usepackage{xcolor}
\usepackage{newfloat}
\usepackage{listings}
\DeclareCaptionStyle{ruled}{labelfont=normalfont,labelsep=colon,strut=off} 
\floatstyle{ruled}
\newfloat{listing}{tb}{lst}{}
\floatname{listing}{Listing}
\title{SECap: Speech Emotion Captioning with Large Language Model}

\author {
    Yaoxun Xu\textsuperscript{\rm 1},
    Hangting Chen\textsuperscript{\rm 2},
    Jianwei Yu\textsuperscript{\rm 2,*},
    Qiaochu Huang\textsuperscript{\rm 1},
    \\Zhiyong Wu\textsuperscript{\rm 1,\rm 3,\thanks{Corresponding authors.}}, 
    Shixiong Zhang\textsuperscript{\rm 2}, 
    Guangzhi Li\textsuperscript{\rm 2}, 
    Yi Luo\textsuperscript{\rm 2}, 
    Rongzhi Gu\textsuperscript{\rm 2}
}
\affiliations{
    \textsuperscript{\rm 1}Shenzhen International Graduate School, Tsinghua University, Shenzhen, China\\

    \textsuperscript{\rm 2}Tencent AI Lab
    \textsuperscript{\rm 3}The Chinese University of Hong Kong, Hong Kong SAR, China\\

    xuyx22$@$mails.tsinghua.edu.cn, \{erichtchen,tomasyu\}$@$tencent.com, zywu$@$sz.tsinghua.edu.cn
}

\begin{document}

\maketitle

\begin{abstract}
Speech emotions are crucial in human communication and are extensively used in fields like speech synthesis and natural language understanding. Most prior studies, such as speech emotion recognition, have categorized speech emotions into a fixed set of classes.  Yet, emotions expressed in human speech are often complex, and categorizing them into predefined groups can be insufficient to adequately represent speech emotions. On the contrary, describing speech emotions directly by means of natural language may be a more effective approach. Regrettably, there are not many studies available that have focused on this direction.
Therefore, this paper proposes a speech emotion captioning framework named \textit{SECap},
aiming at effectively describing speech emotions using natural language.
Owing to the impressive capabilities of large language models in language comprehension and text generation, SECap employs LLaMA as the text decoder to allow the production of coherent speech emotion captions. In addition, SECap leverages HuBERT as the audio encoder to extract general speech features and Q-Former as the Bridge-Net to provide LLaMA with emotion-related speech features. To accomplish this, Q-Former utilizes mutual information learning to disentangle emotion-related speech features and speech contents, while implementing contrastive learning to extract more emotion-related speech features. 
The results of objective and subjective evaluations demonstrate that:
1) the SECap framework outperforms the HTSAT-BART baseline in all objective evaluations;
2) SECap can generate high-quality speech emotion captions that attain performance on par with human annotators in subjective mean opinion score tests. 


\end{abstract}

\section{Introduction}
Speech communication plays a pivotal role in people's daily life in terms of transmitting information and establishing connections. As one of the core carriers of interpersonal communication, speech not only undertakes the function of verbal communication but also deeply involves the transmission of emotions and intentions.  Recognizing and interpreting speech emotions precisely is crucial for enhancing communication effects. Therefore, how to extract the speaker's emotional information accurately from speech has gradually become an important topic in the field of speech processing.


Previous research has typically approached speech emotion acquisition as a categorization task, known as speech emotion recognition (SER) \cite{ser,ser1,ser2}, where emotions like fear and happiness are assigned to discrete categories. In recent years, the performance of such SER tasks has made great progress thanks to the emergence of innovative model architectures.

However, traditional SER exhibits limitations, because single-word labels often lack nuances, failing to convey detailed emotional information like intensity and fluctuations. Speech emotions are typically multifaceted, encompassing diverse affective states (e.g., simultaneous happiness and nervousness) within one utterance. Classifying speech into a single emotion category may inadequately capture authentic emotion. Additionally, the inherently subjective perception of emotions leads to potential variability in emotion classification among individuals interpreting complicated speech.
Considering the limitations of speech emotion classification, employing natural language sentences rather than labels could be a promising strategy to describe speech emotions more precisely. 
Motivated by the recent progress of the Automated Audio Captioning (AAC) task \cite{aacintro,aacintro1,maac} which employs natural language to describe acoustic events in audio, we present the Speech Emotion Captioning (SEC) task and propose an innovative SECap framework, comprising an audio encoder, a Bridge-Net, and a text decoder, to characterize human speech emotions using natural language.  To our knowledge, this is among the pioneering works in this direction.

In the SEC task, there are two primary challenges to address: firstly, how to extract the emotion-related speech features from the original speech inputs; and secondly, how to generate high-quality, human-like speech emotion descriptions. 
For the first challenge, limited speech data with emotion captions makes training the audio encoder from scratch challenging. Inspired by the success of pre-trained model in SER \cite{hubertser} tasks, we utilize HuBERT \cite{hsu2021hubert} as SECap's audio encoder for robust speech feature extraction. 
However, directly using the frame-level HuBERT features can be computationally heavy. To address this, inspired by BLIP-2 \cite{li2023blip}, we employ Q-Former as Bridge-Net to compress HuBERT features. 
While both acoustic and content information within HuBERT features are related to speech emotion, acoustic information is typically more directly related to speech emotion, and content information can be easily obtained through transcription. Therefore, in the Bridge-Net,  we aim to separately  extract emotion-related acoustic information from HuBERT features while eliminating content information. Thus, we employ Speech-Caption Contrastive Learning and Speech-Transcription Mutual Information Learning to train Bridge-Net to better extract emotion-related acoustic information.

For the second challenge, due to the advances in large language models (LLMs) and their impressive natural language understanding capabilities, such as GPT-4 \cite{gpt4}, we employ LLaMA \cite{touvron2023llama} as text decoder for generating fluent and coherent speech emotion captions based on Q-Former-extracted speech features. Concurrently, we use LLaMA to guide Q-Former training, enabling better projection of speech emotion features into LLaMA, ultimately yielding higher-quality speech emotion captions.

As for evaluation, we design both subjective and objective evaluation metrics based on the AAC task to better assess the quality of speech emotion captions generated by SECap. To facilitate a more effective comparison, we choose the HTSAT-BART model \cite{mei2023wavcaps}, which performs exceptionally well in the AAC task, as our baseline. Experimental results demonstrate that SECap outperforms the HTSAT-BART model across all objective metrics. In the subjective mean opinion score (MOS) test, the quality of speech emotion captions generated by SECap surpasses that of human labels (i.e., 3.77 vs. 3.39 MOS score) and are on par with human annotations (i.e., 3.77 vs. 3.85 MOS score). Our main contributions are as follows:

\begin{itemize}
    \item We propose the task of Speech Emotion Captioning (SEC), which, to our knowledge, stands among the pioneering efforts to characterize speech emotions using natural language.
    \item We introduce SECap\footnote{Codes, models and results: https://github.com/thuhcsi/SECap} to tackle the SEC task, which comprises a HuBERT-based audio encoder, a Q-Former-based Bridge-Net, and a LLaMA-based text decoder. 
    \item Experimental results show that SECap is capable of generating suitable and fluent speech emotion captions that are on par with human-labeled speech emotion captions.
    
\end{itemize}

\begin{figure}[t]
\centering
\includegraphics[width=0.95\columnwidth]{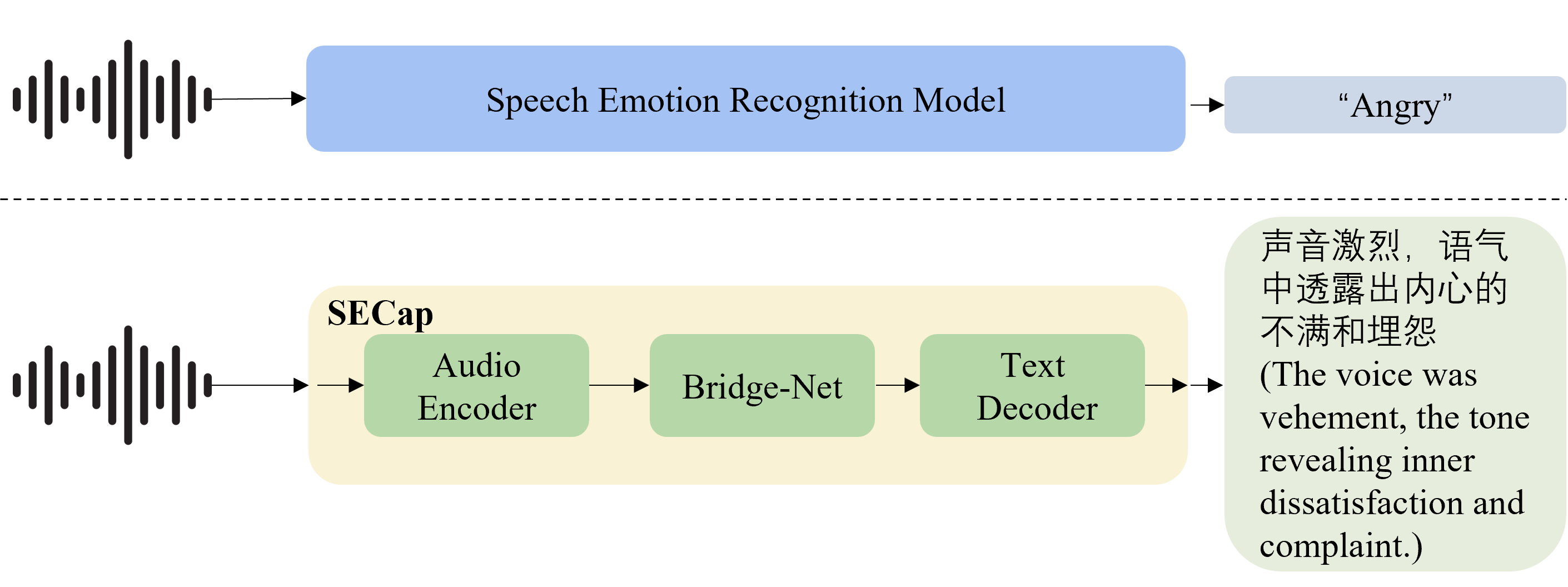} 
\caption{Comparison of Speech Emotion Recognition (SER) model and the proposed SECap. The SER model generates emotion labels, while SECap generates natural language emotion descriptions derived from the speech. }
\label{fig1}
\end{figure}

\begin{figure*}[th]
\centering
\includegraphics[width=1.6\columnwidth]{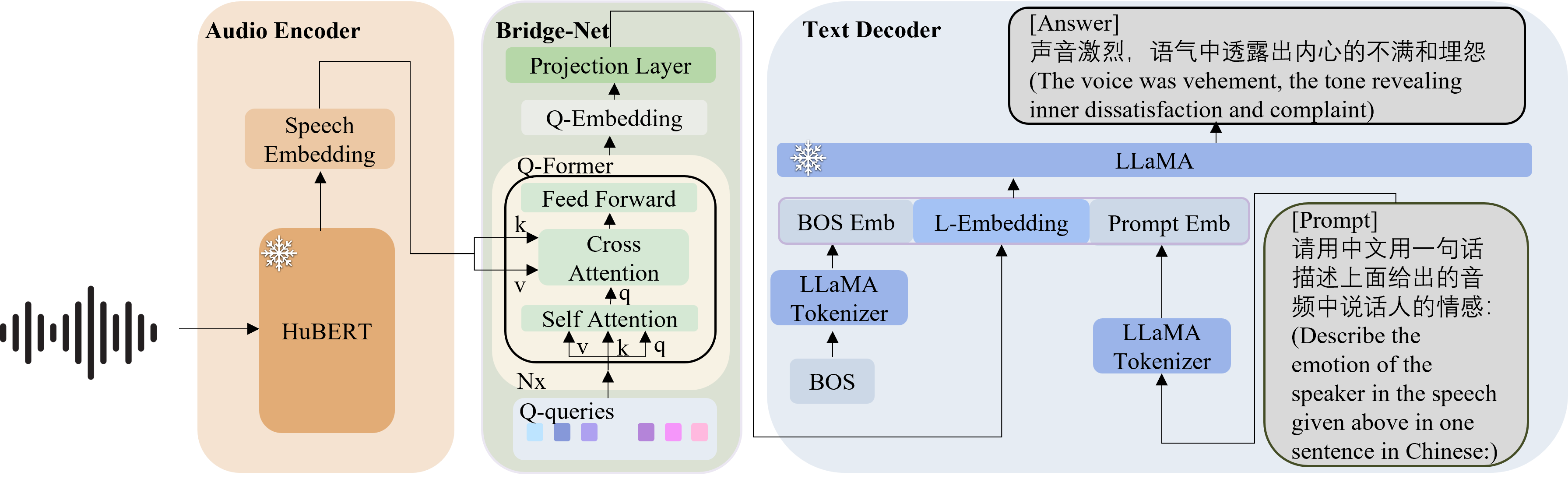} 
\caption{Framework of the proposed SECap}
\label{fig1}
\end{figure*}
\section{Related Work}
\subsection{Speech Emotion Recognition}
Speech Emotion Recognition (SER) entails detecting and classifying emotions in spoken language, ultimately categorizing them into specific labels. From the perspective of pattern recognition, SER \cite{serreview} can be divided into three components: feature extraction, feature selection, and feature classification. Extracted features include Mel-Frequency Cepstral Coefficients (MFCC) \cite{mfcc}, and Linear Predictive Coding (LPC) \cite{lpc}, among others. Owing to the advancement of deep learning, feature classifiers have evolved from methods like Linear Discriminant Analysis (LDA) \cite{lda} to neural network architectures such as CNN and Transformer \cite{vaswani2017attention}. Our approach describes speech emotions with natural sentences rather than confining them to specific categories.
\subsection{Large Language Model}
Large Language Models (LLMs) \cite{llm1,llm2,llm3} have revolutionized natural language understanding.
By analyzing vast textual data, LLMs learn linguistic patterns and generate natural prose. Open-source models like BLOOM \cite{scao2022bloom} and ChatGLM \cite{du2022glm} have fostered growth in the LLM community. Also, researchers explore LLMs' performance in multimodal interactions, aspiring for models capable of managing audio, vision, and text modalities, reflecting human daily interactions. The first approach uses LLMs as task orchestrators, connecting downstream models like AudioGPT \cite{huang2023audiogpt} and HuggingGPT \cite{shen2023hugginggpt} for specialized tasks. The second approach positions LLMs as multitask processors, mapping modal tasks to a unified space. For example, BLIP-2 maps images to text space using Q-Former, while Video-LLaMA \cite{videollama} maps audio and vision modalities via Q-Former. 

\subsection{Automated Audio Captioning}
Automated audio captioning (AAC) \cite{aac,aacrelate} is a crucial task in the audio domain, describing ambient sounds using natural language. Unlike audio tagging \cite{audiotag} or sound event detection \cite{sed}, AAC requires identifying specific events and describing them naturally. After the emergence of this task, the encoder-decoder \cite{s2s} framework has been the dominant solution to this problem. Methods like AudioClip \cite{guzhov2022audioclip} and CLAP \cite{clap} 
employ contrastive learning to map audio and text, boosting the encoder-decoder connection. 

\section{Method}
Inspired by the AAC task, SECap employs encoder-decoder architecture, as illustrated in Figure 1. The audio encoder extracts speech features, and Bridge-Net extracts emotion-related speech features and transforms them into the text decoder's feature space. The text decoder then generates speech emotion captions based on these features.

In this section, we will begin by providing an overview of the SECap structure. Following that, we will elaborate on the two key aspects of the Bridge-Net design for obtaining emotion-related representations. Lastly, we will describe the overall training process of SECap.

\subsection{Model Architecture}

As illustrated in Figure 2, SECap utilizes a HuBERT-based audio encoder, a Q-Former-based Bridge-Net, and a LLaMA-based text decoder. 

The HuBERT is to derive speech embedding for its powerful speech feature extraction capability. However, frame-level HuBERT features can lead to heavy computation cost. We employ Q-Former-based Bridge-Net to compress features. Meanwhile, acoustic information is more directly associated with speech emotion, while content information is obtainable from transcriptions. Thus, the Bridge-Net is used to extract emotion-related acoustic information and eliminate content information. We employ LLaMA as the text decoder for generating speech emotion captions, leveraging its exceptional language comprehension capabilities. Aligning with LLaMA's input format, we position L-Embedding between the ``BOS" and a prompt. This method constrains LLaMA's output space via the prompt, yielding more accurate speech emotion captions.

\subsubsection{Q-Former}

Owing to the redundancy of HuBERT speech features, Q-Former is designed and adopted to compress and extract emotion-related speech features which consists of self-attention, cross-attention, and linear layers. Q-queries are learnable parameters for extracting speech embedding. 

Let $q \in \mathbb{R}^{n_q \times d_q}$ represent the Q-queries, where $n_q$ is the number of Q-queries, $d_q$ is the dimension of Q-queries and $S \in \mathbb{R}^{n_s \times T_s \times d_s}$ represent the speech embedding, where $n_s$ is the batch size, $T_s$ is the number of time steps, and $d_s$ is the dimension of speech embedding.
We first input the Q-queries $q \in \mathbb{R}^{n_q \times d_q}$ into the self-attention mechanism:
\begin{equation}
A_{\text{self}} = \text{softmax} \left( \frac{q W_{q_{\text{self}}} (q W_{k_{\text{self}}})^T}{\sqrt{d_k}} \right) q W_{v_{\text{self}}}
\end{equation}
where $W_{q_{\text{self}}} \in \mathbb{R}^{d_q \times d_k}$, $W_{k_{\text{self}}} \in \mathbb{R}^{d_q \times d_k}$, and $W_{v_{\text{self}}} \in \mathbb{R}^{d_q \times d_v}$ are the learnable weight matrices for queries, keys, and values in the self-attention mechanism, and $d_k$ and $d_v$ are the dimensions of keys and values.
The output of the self-attention mechanism $A_{\text{self}} \in \mathbb{R}^{n_q \times d_v}$ are then used as the queries for the cross-attention mechanism, while the speech embedding $S \in \mathbb{R}^{n_s \times T_s \times d_s}$ serve as keys and values:
\begin{equation}
A_{\text{cross}} = \text{softmax} \left( \frac{A_{\text{self}} W_q (S W_k)^T}{\sqrt{d_k}} \right) S W_v
\end{equation}
where $A_{\text{cross}} \in \mathbb{R}^{n_s \times n_q \times d_v}$ represents the cross-attention output,
while $W_q \in \mathbb{R}^{d_v \times d_k}$, $W_k \in \mathbb{R}^{d_s \times d_k}$, and $W_v \in \mathbb{R}^{d_s \times d_v}$ are the learnable weight matrices for queries, keys, and values in the cross-attention mechanism.

This approach enables the attention mechanism to retrieve features related to Q-queries within the speech embedding. Specifically, the output of the Q-Former, denoted as the Q-Embedding $Q_e \in \mathbb{R}^{n_s \times n_q \times d_q}$, maintains a fixed length that is independent of the length of the input speech. 
This fixed-length representation leads to improved generalization performance across speech inputs of varying lengths.

\begin{figure*}[t]
\centering
\includegraphics[width=1.6\columnwidth]{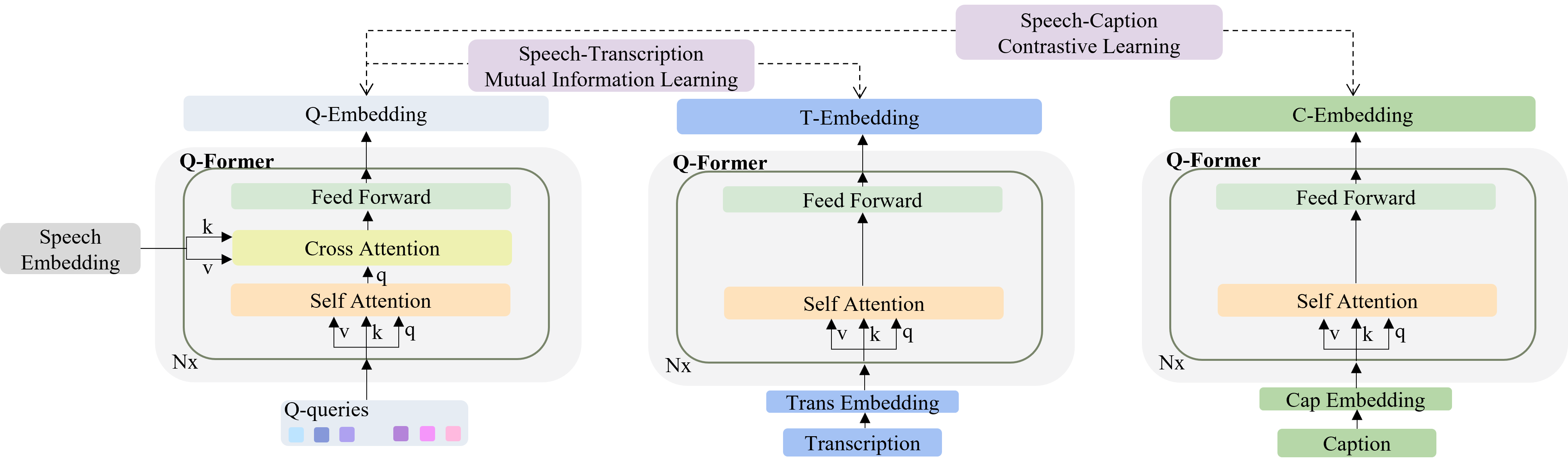} 
\caption{The figure presents Q-Former decoupling audio representation and content information using Speech-Transcription Mutual Information Learning with speech features (Q-Embedding) and speech transcription features (T-Embedding). Additionally, it obtains more emotion-related audio representation through Speech-Caption Contrastive Learning with speech features (Q-Embedding) and speech emotion caption features (C-Embedding).}
\label{fig1}
\end{figure*}

\subsection{Obtain Emotion-Related Representations}
\label{ssec:subhead}
To provide LLaMA with more content-unrelated and emotion-related speech features, we simultaneously incorporate both the human-labeled speech emotion captions and the transcriptions. As depicted in Figure 3, these are passed through a Q-Former that is largely consistent with the original, except for the absence of the cross-attention module. This process yields C-Embedding $Q_c \in \mathbb{R}^{n_s \times T_c \times d_q}$ and T-Embedding $Q_t \in \mathbb{R}^{n_s \times T_t \times d_q}$, where $T_c$ and $T_t$ denote the length of the caption and the transcription, respectively. We employ Speech-Transcription Mutual Information Learning to disentangle speech features from speech content. Additionally, Speech-Caption Contrastive Learning  is utilized to extract more emotion-related speech features. 

\subsubsection{Speech-Transcription Mutual Information Learning (STMIL)}


The speech content can potentially impact emotion assessment, for example, expressing joyous statements calmly.
To minimize the correlation between speech features and content, thereby mitigating the speech content's impact on LLaMA's speech emotion caption generation, we propose Speech-Transcription Mutual Information Learning. As illustrated in Figure 3, we introduce both Speech Embedding and Trans Embedding into the Q-Former simultaneously, yielding Q-Embedding $Q_e$  and T-Embedding $Q_t$. This enables the comparison of speech and its content within a unified representation space. To evaluate the correlation between $Q_e$ and $Q_t$, we adopt mutual information $I(Q_t; Q_e)$ as the metric:
\begin{equation}
I(Q_t; Q_e) = \sum_{q_t \in Q_t} \sum_{q_e \in Q_e} p(q_t, q_e) \log \frac{p(q_t, q_e)}{p(q_t)p(q_e)}
\end{equation}
where $p(q_t, q_e)$ represents the joint probability distribution of $Q_t$ and $Q_e$, and $p(q_t)$ and $p(q_e)$ denote the marginal probability distributions of $Q_t$ and $Q_e$, respectively.

However, direct computation of mutual information between $Q_e$ and $Q_t$ is infeasible due to their unknown, high-dimensional nature. While prior methods such as MINE \cite{belghazi2018mine} and infoNCE \cite{van2017infonce} can estimate the lower bound of mutual information,  they are not suitable for controlling the minimization process.
Following vCLUB \cite{cheng2020club}, we use Equation (4) to estimate the upper bound of mutual information and employ it as a loss function to reduce the correlation between speech features and content. 
\begin{equation}
\mathcal{U}(Q_t; Q_e) = \frac{1}{n^2} \sum_{i=1}^n \sum_{j=1}^n \left[ \log \frac{q(y_i | x_i)}{q(y_j | x_i)} \right]
\end{equation}
The equation includes conditional probabilities $q(y_i | x_i)$ and $q(y_j | x_i)$, representing the probabilities of the $i$-th and $j$-th $Q_e$ samples given the $i$-th $Q_t$ sample. The logarithm captures the dissimilarity between $Q_e$ conditioned on $Q_t$, and summing over all pairwise combinations provides the upper bound mutual information measure between $Q_e$ and $Q_t$.
 
\subsubsection{Speech-Caption Contrastive Learning (SCCL)}
\label{ssec:subhead}

Due to the high dimensionality and redundancy in speech representations, speech features contain abundant information such as content and background noise, with only a fraction being emotion-related. To alleviate the complexity of processing speech features by LLaMA, we aim for Q-Former to extract features highly correlated with speech emotion caption, consequently bridging the gap between speech features and text modality. As illustrated in Figure 3, our objective is to minimize the distance between $Q_e$ and C-Embedding $Q_c$, prompting Q-Former to extract more emotion-related features and progressively approach the text modality. Drawing inspiration from CLAP \cite{clap}, we employ a contrastive learning approach to accurately represent distances between $Q_e$ of distinct speech samples, ensuring that speech with similar emotions yield closer $Q_e$ distances, while those with dissimilar emotions result in farther $Q_e$ distances.

To mitigate the influence of similar emotions in negative samples during contrastive learning, we partition the dataset into $N$ distinct categories based on human-labeled speech emotion labels. This guarantees substantial differences in speech emotion captions across categories, thereby enhancing the model's discriminative capacity throughout the learning process. 
During each training step, we select $K$ speech-caption pairs from each of the $N$ sets, ensuring that for each $Q_e$ (referred to as $e_i$), there is 1 corresponding $Q_c$ (referred to as $d_i$),$(K-1)$ $Q_c$ with similar emotions (referred to as $p_i$), and $(NK-K)$ $Q_c$ with dissimilar emotions (referred to as $u_i$).

We opt to use cosine similarity $\mathcal{S}$ to measure the distance between $Q_e$ and $Q_c$. For enhanced contrastive learning, we design the training method as follows:
\begin{align}
\mathcal{L}(Q_c;Q_e) = \sum_{i=1}^{NK}\Big[ w_1 (1-\mathcal{S}(e_i, d_i)) + w_2 \sum_{j=1}^{K-1} (1- \nonumber \\
\begin{aligned}
&\mathcal{S}(e_i, p_{ij})) + w_3 \sum_{j=1}^{NK - K} \text{ReLU}( \mathcal{S}(e_i, u_{ij}) - m) \Big]
\end{aligned}
\end{align}
where the weighting coefficients $w_1$, $w_2$, and $w_3$ control the contribution of each term in the loss function. The threshold value $m$ is the margin to control the distance between speech feature $Q_e$ and irrelevant speech emotion caption feature $Q_c$.


\subsection{Training Process}
\label{ssec:subhead}
To enhance speech emotion caption generation with LLaMA, we devise a two-stage training process. 
The first stage compresses HuBERT-extracted speech features to obtain emotion-relevant attributes, while the subsequent stage aligns these features with LLaMA's representation space.

In the first training stage, we combine STMIL and SCCL for collaborative training as in Figure 3, while keeping the HuBERT model frozen.
Inspired by BLIP-2, we initialize the Q-Former using pre-trained parameters from $\text{BERT}_\text{base}$ \cite{bert}. 
Specifically, the training loss is:
\begin{equation}
 \mathcal{L}_{T1} = w_{T1} \times \mathcal{U}(Q_t; Q_e) + w_{T2} \times \mathcal{L}(Q_c;Q_e)
\end{equation}
where the weighting coefficients $w_{T1}$ and $w_{T2}$ control the contribution of STMIL and SCCL. 

In the second training stage, we fine-tune the Q-Former and the projection layer to effectively integrate Q-Former-extracted speech features into LLaMA.
Meanwhile, the parameters of LLaMA and HuBERT remain frozen.
We insert a ``BOS" token before the L-Embedding to align with the inference format. To improve SECap's generalization ability, we devise 30 semantically akin sentences, each instructing to ``portray the speaker's emotion in a single Chinese sentence." During training, we randomly choose a sentence to concatenate after the L-Embedding. Subsequently, we append the human-labeled speech caption $C$ after the prompt and employ the teacher-forcing approach to enable LLaMA generate caption $\hat{C}$. 
Cross-entropy loss (CELoss) is then adopted as the training objective:
\begin{equation}
\mathcal{L}_{T2} = \text{CELoss}(C,\hat{C})
\end{equation}

\section{Dataset}
Due to the lack of publicly available SEC datasets, we utilize an internal dataset called EMOSpeech. EMOSpeech dataset consists of 5 female and 2 male speakers, totaling 41.6 hours of speech covering 30526 sentences, sampled at a rate of 24kHz. Each speech in EMOSpeech has three to five human-labeled speech emotion captions and human-labeled speech emotion labels provided by different annotators, along with its corresponding transcription.



As for labeling, we begin with 50 sample audio clips for independent annotator labeling and hold a discussion session for annotators to review annotations and establish standardized rules based on collective input.
The annotation process has three levels: identifying overall emotion using a single word, describing emotion intensity, and providing a comprehensive sentence considering emotion, volume, and speech rate. 
With these guidelines, annotators consistently label the dataset. To ensure annotation quality, we conduct consistency checks by randomly selecting 5 out of every 100 clips for review by other annotators, upholding high standards throughout the dataset construction.

Upon constructing the EMOSpeech dataset, we randomly select 600 sentences for testing, 600 sentences for validation, and the remaining 29,326 sentences for training\footnote{Please refer to project's GitHub repository for detailed dataset construction process, where the test set is also publicly available.}.

\section{Evaluation Metric}

Since there is no existing method to evaluate speech emotion captions, we devise both objective and subjective evaluation methods based on the nature of the SEC task. 
\subsection{Objective Evaluation}
In this study, we initially adopt objective evaluation metrics for the AAC task, containing BLEU$_1$ \cite{bleu}, BLEU$_4$, METEOR \cite{meteor}, ROUGE$_l$ \cite{lin2004rouge},  CIDEr \cite{cider}, and SPICE \cite{anderson2016spice}.
However, these metrics primarily focus on word-level matching.
To more effectively assess the similarity between two Chinese emotion captions at the sentence level, 
we employ sentence similarity evaluation metrics in conjunction with the aforementioned criteria. The first model \cite{text2vec} is based on MACBERT \cite{cui2021macbert} and trained on Chinese STS-B \cite{sts}, while the second model \cite{reimers2019sentence} is finetuned on Tencent Cloud. Their evaluation indicators are denoted as SIM$_1$ and SIM$_2$, respectively.

\subsection{Subjective Evaluation}
In the subjective scoring method, we develop a three-stage scoring criterion to reduce variability due to evaluators' inconsistent understanding of emotions. The first step involves determining whether the generated sentence describes an emotion. The second step assesses if the generated sentence, when summarized into an emotion, matches the speech. The third step evaluates whether the generated sentence aligns with the speech in terms of the emotion's intensity.

To be specific, we have devised a scoring method similar to the Mean Opinion Score (MOS) used in Text-to-Speech systems, with ratings ranging from 1 to 5, where 1 represents the worst and 5 represents the best.


\section{Results and Analysis}
\subsection{Experiment Setup}
Our experiments are conducted exclusively on the EMOSpeech dataset. We choose the HuBERT-large model, pre-trained on the 10k-hour WenetSpeech \cite{zhang2022wenetspeech} L subset, as the audio encoder. Due to the original LLaMA's limited proficiency in understanding Chinese, we choose an enhanced version of LLaMA \cite{cui2023llama11}  finetuned with Chinese datasets as the text decoder\footnote{Specific experimental details are given at GitHub repository.}.
\begin{table}[H]
\caption{The number of trainable parameters and the total parameters during the two stages of the training process}
\centering
\small
\begin{tabular}{ccc}
\toprule[2pt]
                 & Trainable Params & Total Params \\ \hline
Training stage 1 & 100M             & 517M         \\
Training stage 2 & 103M             & 7.4B       \\
\bottomrule[2pt]
\end{tabular}

\end{table}

\begin{table*}[t]
\caption{Objective experiment results on performance analysis. ``Raw Trans'' denotes raw transcription of speech.``Q-Emb'' denotes Q-Embedding.``T-Emb'' denotes T-Embedding. ``\textemdash" refers to not included. Higher scores mean better performance.}
\centering
\small
\begin{tabular}{c|c|cc|cc|cccccc}
\toprule[2pt]
\multirow{2}{*}{Model} & \multirow{2}{*}{ID} & \multicolumn{2}{c|}{Input Modality}                     & \multirow{2}{*}{SIM$_1$}        & \multirow{2}{*}{SIM$_2$}        & \multirow{2}{*}{BLEU$_1$}       & \multirow{2}{*}{BLUE$_4$}      & \multirow{2}{*}{METEOR}         & \multirow{2}{*}{ROUGE$_l$}      & \multirow{2}{*}{CIDEr}          & \multirow{2}{*}{SPICE}         \\ \cline{3-4}
                       &                     & Text                       & Audio                      &                                 &                                 &                                 &                                &                                 &                                 &                                 &                                \\ \hline
HTSAT-BART             & \#1                 & \textemdash & Q-Emb                      & 59.62                           & 53.19                           & 32.74                           & 3.05                           & 14.61                           & 23.64                           & 2.21                            & 2.17                           \\ \hline
\multirow{5}{*}{SECap} & \#2                 & Raw Trans                  & \textemdash & 56.49                           & 22.38                           & 0.014                           & 0.00                           & 2.67                            & 4.38                            & 0.00                            & 0.00                           \\
                       & \#3                 & T-Emb                      & \textemdash & 65.90                           & 62.24                           & 25.97                           & 4.28                           & 15.98                           & 23.37                           & 18.37                           & 2.58                           \\
                       & \#4                 & Raw Trans                  & Q-Emb                      & 69.24                           & 67.50                           & 29.59                           & 5.36                           & 16.99                           & 25.16                           & 28.51                           & 5.77                           \\
                       & \#5                 & T-Emb                      & Q-Emb                      & 69.66                           & 70.02                           & 33.62                           & 7.25                           & 18.44                           & 27.18                           & 33.82                           & 5.96                           \\
                       & \#6                 & \textemdash & Q-Emb                      & \textbf{71.95} & \textbf{70.51} & \textbf{36.08} & \textbf{8.12} & \textbf{19.30} & \textbf{28.49} & \textbf{34.81} & \textbf{6.49}\\
\bottomrule[2pt]
\end{tabular}

\end{table*}

\subsection{Performance Analysis}


This experiment aims to demonstrate the effectiveness of SECap. Given speech content's impact on emotion, we incorporate transcriptions as additional input and design multiple comparison groups for thorough analysis. Specifically, when incorporating transcriptions, we either use the raw transcriptions processed through the LLaMA tokenizer or apply the T-Embedding processed through a projection layer as an extra input, which is then concatenated between BOS Emb and L-Embedding. Subsequently, we append the Prompt Emb and feed it into LLaMA. 
In some of the comparison groups, speech features are not used, the former processed transcriptions are directly concatenated between BOS Emb and Prompt Emb.
%
\subsubsection{Objective Evaluation}

Table 2 illustrates consistent trends in most metrics across various experiments. Therefore, our analysis will primarily focus on the insights obtained from the two Chinese sentence similarity models.

As depicted in Table 2, when incorporating only speech features, the proposed SECap (\#6) surpasses the HTSAT-BART (\#1) baseline across all objective metrics, signifying its ability to generate more natural, human-like speech emotion captions compared to the HTSAT-BART model.

Compared to raw transcription (\#2), employing T-Embedding  (\#3) improves SIM values by 16.66\% and 178.11\%. Since LLaMA has not been previously trained on the EMOSpeech dataset, it lacks prior knowledge of the dataset's captions, resulting in unconstrained output space. However, T-Embedding imposes greater constraints, extracting more emotion-related features, and resulting in relatively accurate speech emotion captions.


Using only Q-Embedding (\#6), we observe a 9.18\% and 13.29\% SIM value increase compared to relying solely on T-Embedding (\#3). As identical sentences can convey different emotions, relying only on speech content 
(i.e. transcription) 
may not sufficiently reflect speech emotions, while 
speech signals
better represent speech emotions. Upon incorporating Q-Embedding and the raw transcription (\#4), the SIM values relatively decrease by 3.77\% and 4.27\% compared to only using Q-Embedding (\#6). However, replacing the raw transcription (\#4) with T-Embedding (\#5) shows a relative improvement of 0.61\% and 3.73\% in SIM values. Despite this increase, the SIM values remain lower than those obtained when using only Q-Embedding (\#6).




Consistent with text-only modality, T-Embedding outperforms raw transcription in extracting emotion features from transcription, providing greater constraints for LLaMA while reducing conflicts with Q-Embedding. However, integrating both audio and text modalities into the model may increase the difficulty for LLaMA in processing the information, as text and audio could contain similar, unrelated, or even contradictory information. Consequently, LLaMA must balance these features, potentially impeding its ability to optimally leverage information from both modalities and affecting the model's assessment of speech emotion.

\subsubsection{Subjective Evaluation}
In the subjective experiment, we randomly select a test set of 50 sentences. We apply all the methods listed in Table 2 to generate corresponding speech emotion captions. To enable a more comprehensive comparison, we also include human-labeled speech emotion labels (Human Label) and emotion categories (SER Model Label)  identified by a competitive pre-trained Chinese SER model from HuggingFace\footnote{https://huggingface.co/xmj2002/hubert-base-ch-speech-emotion-recognition}. We request evaluators to score the nine types of texts according to the Subjective Evaluation details provided in Evaluation Metric section. 
15 evaluators participate in the tests, with the results presented in Figure 4.
\begin{figure}[h]
\centering
\includegraphics[width=1.0\columnwidth]{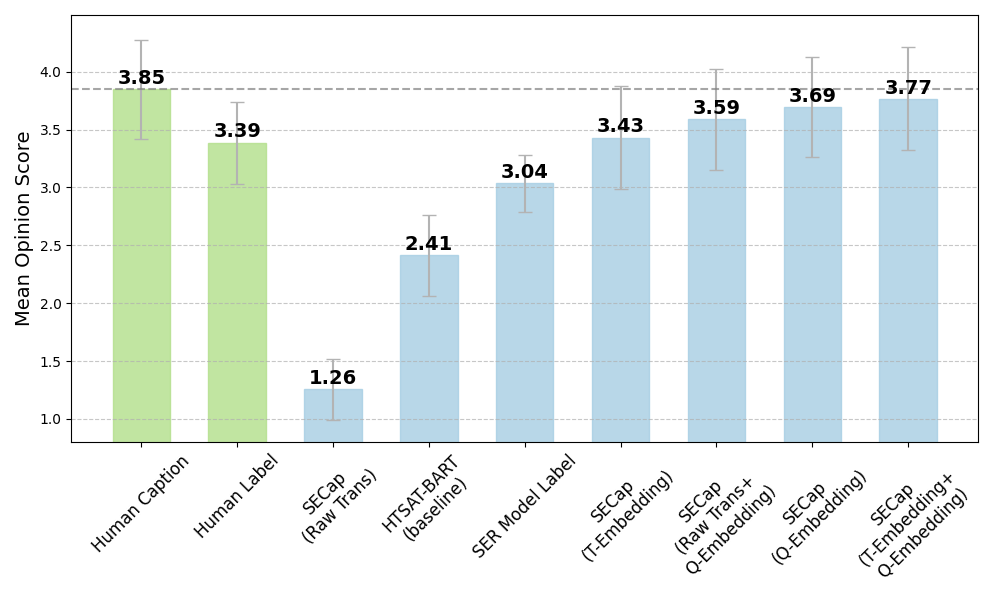} 
\caption{Subjective experiment results on performance analysis}
\label{fig1}
\end{figure}

As seen in Figure 4, human-labeled speech emotion captions surpass both human-labeled speech emotion labels and SER Model Labels. This outcome aligns with the SEC task's aim to represent emotions more comprehensively and accurately within a single sentence. Notably, the best SECap model outperforms human-labeled speech emotion labels and is on par with human-labeled speech emotion captions.

Furthermore, Figure 4 reveals that SECap's performance is suboptimal with solely raw transcription. However, other SECap input methods outperform human-labeled emotion labels and the baseline in subjective evaluation metrics, demonstrating SECap's ability to generate suitable speech emotion captions deemed more representative of emotions. However, employing both Q-Embedding and T-Embedding as input generates superior subjective evaluation outcomes compared to using solely Q-Embedding.


We suppose that objective metrics, relying on predefined rules, may differ from subjective evaluations based on human perception and understanding. Discrepancies can arise as evaluators focus on intricate details, such as emotional expression's naturalness and contextual information, which are challenging for objective metrics to capture. Additionally, we observed that evaluators initially concentrated on speech content. In cases of content-emotion conflicts, evaluators tended to assign higher scores to content-relevant captions. For example, uttering ``I'm feeling terrible today" in a flat tone with only speech embedding might yield a caption describing the flat tone, while incorporating text embedding could result in a caption combining sadness and flatness.

\subsection{Ablation Study on Different Model Components}

This experiment aims to explore the effect of different model components on generating speech emotion captions. Considering that HTSAT-BART differs from SECap in both the audio encoder, the Bridge-Net, and the text decoder, to better analyze each component, we construct the model using different audio encoders, text decoders, and Bridge-Nets, respectively. In the previous experiment, we find that the two text similarity models exhibit the same trend despite different experiments. Therefore, in this experiment as well as subsequent experiments, we only use the first text similarity model introduced in the Evaluation Metric section for the evaluation of objective indicators.
\begin{table}[h]
\caption{Experiment results on the ablation study of different model components}
\centering
\small
\begin{tabular}{ccc|c}
\toprule[2pt]
\multicolumn{3}{c|}{Model}                & \multirow{2}{*}{SIM$_1$}    \\ \cline{1-3}
Audio Encoder & Bridge-Net & Text Decoder &                         \\ \hline
HTSAT         & Linear     & BART         & 59.62$\pm$0.22          \\
HuBERT        & Linear     & BART         & 63.95$\pm$0.27          \\
HTSAT         & Linear     & LLaMA        & 64.94$\pm$0.06          \\
HuBERT        & Linear     & LLaMA        & 68.62$\pm$0.32          \\
HuBERT        & Q-Former   & LLaMA        & \textbf{71.95$\pm$0.04}\\
\bottomrule[2pt]
\end{tabular}

\end{table}

Table 3 reveals that by replacing the text decoder with LLaMA and retaining the audio encoder, the HTSAT-BART's SIM value increases by 8.92\%, indicating LLaMA's superior text generation capabilities compared to BART. Similarly, substituting the audio encoder with HuBERT while maintaining the text decoder results in a 7.26\% SIM value increase, suggesting HuBERT's robust speech feature extraction abilities compared to HTSAT. Replacing both components yields a 15.10\% SIM value increase. Furthermore, exchanging the linear layer with Q-Former significantly enhances high-quality speech emotion caption generation, accompanied by a 4.85\% SIM value increase.


Evidently, HuBERT is more suitable for speech feature extraction compared to HTSAT, and LLaMA demonstrates superior text comprehension and generation capabilities compared to BART. By further extracting speech features using the Q-Former, speech features that better align with emotional aspects can be conveyed to LLaMA, ultimately resulting in more accurate speech emotion captions. 

\subsection{Comparison of Training Methods}
This experiment seeks to investigate the impact of various training methods for the Q-Former on generating speech emotion captions. While maintaining the audio encoder as HuBERT and the text decoder as LLaMA, we conduct a series of comparisons, including whether to use STMIL or SCCL in the first training stage, and whether to freeze the Q-Former in the second training stage.

\begin{table}[h]
\caption{Experiment results on the effect of different methods of training the Q-Former}
\centering
\small
\begin{tabular}{ccccc}
\toprule[2pt]
                     & \multicolumn{3}{c}{Method}                                                              & \multirow{2}{*}{SIM$_1$}                     \\ \cline{2-4}
                     & STMIL                         & SCCL                         & Freeze                      &                                          \\ \hline
HTSAT-BART           & \XSolidBrush & \XSolidBrush & \XSolidBrush & 63.95$\pm$0.27                           \\ \hline
\multirow{5}{*}{SECap}                & \XSolidBrush & \XSolidBrush & \XSolidBrush & 67.29$\pm$0.22                           \\
                     & \Checkmark   & \XSolidBrush & \XSolidBrush & 68.75$\pm$0.12                           \\
                     & \XSolidBrush & \Checkmark   & \XSolidBrush & 69.40$\pm$0.11                           \\
\multicolumn{1}{l}{} & \Checkmark   & \Checkmark   & \XSolidBrush & \textbf{71.95$\pm$0.04} \\
                     & \Checkmark   & \Checkmark   & \Checkmark   & 57.92$\pm$0.19   \\
\bottomrule[2pt]
\end{tabular}

\end{table}

Table 4 indicates that incorporating STMIL or SCCL individually during the first training stage results in a 2.17\% and 3.14\% SIM value increase, respectively, compared to omitting this stage, suggesting that disentangling content information or extracting extra emotion-related speech features enhances caption quality. Moreover, using both methods simultaneously yields a 6.92\% SIM value increase. Without both STMIL and SCCL, Q-Former lacks speech feature insight, and limited EMOSpeech risks overfitting. Compared to employing either method alone, employing both STMIL and SCCL leads to the rise of SIM values by 4.65\% and 3.67\%, respectively, highlighting that concurrent utilization of both methods bolsters speech feature extraction, generating more precise speech emotion captions.

Upon finalizing the initial training phase, a 19.50\% SIM value reduction is observed by freezing Q-Former and solely training the projection layer, compared to the unfrozen Q-Former. As LLaMA lacks involvement in guiding Q-Former's training, the extracted features might inadequately align with LLaMA's input, and the projection layer's adaptability fails to offset this misalignment.




\section{Conclusion}

To better represent speech emotions, we introduce an innovative task called speech emotion captioning, which uses natural language descriptions rather than singular labels to characterize speech emotions. Our proposed model, SECap, integrates a HuBERT-based audio encoder, a LLaMA-based text decoder, and a Q-Former-based Bridge-Net. The Q-Former effectively disentangles speech features and speech content information through Speech-Transcription Mutual Information Learning, while extracting more emotion-related speech features via Speech-Caption Contrastive Learning. Impressively, SECap generates high-quality speech emotion captions, with its performance on par with human annotators. This pioneering task and method provide a fresh perspective on speech emotion understanding, fostering a more comprehensive approach to analyzing and interpreting speech emotional expressions.

\section{Acknowledgement}
This work is supported by National Natural Science Foundation of China (62076144), Shenzhen Science and Technology Program (WDZC20220816140515001, JCYJ20220818101014030) and Tencent AI Lab Rhino-Bird Focused Research Program (RBFR2023015).
\bibliography{aaai24}
\end{document}